\documentclass[letterpaper,10pt,conference]{ieeeconf}
\usepackage{ifthen}
\newboolean{ieeenotice}
\setboolean{ieeenotice}{false}

\usepackage[utf8]{inputenc}
\usepackage[T1]{fontenc}
\usepackage{microtype}
\usepackage{amsmath}
\usepackage{amssymb}
\usepackage[english]{babel}
\usepackage[autostyle]{csquotes}
\usepackage[dvipsnames]{xcolor}
\usepackage{graphicx}
\graphicspath{{./}{./}{./}}
\usepackage[backgroundcolor=orange]{todonotes}
\usepackage{siunitx}
\usepackage{tikz}
\usetikzlibrary{arrows,automata,calc,fit,patterns,positioning,backgrounds,shadows,shapes,decorations}
\usepackage{relsize}
\usepackage{ctable}
\usepackage{multirow}
\usepackage{setspace}
\usepackage{paralist}
\usepackage{hyperref}
\hypersetup{linkcolor=black,citecolor=black,urlcolor=black,colorlinks=true}

\addto\extrasenglish{%
}

\newcommand{\CopyrightNotice}{\hbox
  {\parbox{\textwidth}{\center \textsf{\small $ $\\[-11cm]
    This is a pre-print of an article accepted for publication in \emph{Proceedings of the 30th IEEE Intelligent Vehicles Symposium (IV)}.
  }}}
}

\IEEEoverridecommandlockouts
\title{\Large\textbf{Parallel Multi-Hypothesis Algorithm for\\Criticality Estimation in Traffic and Collision Avoidance}}
\author{Eduardo Sánchez Morales$^{1}$, Richard Membarth$^{2}$, Andreas Gaull$^{1}$, Philipp Slusallek$^{2}$,\\ Tobias Dirndorfer$^{3}$, Alexander Kammenhuber$^{3}$, Christoph Lauer$^{3}$, and Michael Botsch$^{1}$%
\thanks{$^{1}$Eduardo Sánchez Morales, Michael Botsch, and Andreas Gaull are with the Department of Vehicle Safety, CARISSMA, Technische Hochschule Ingolstadt, Ingolstadt, Germany.}%
\thanks{$^{2}$Richard Membarth and Philipp Slusallek are with the German Research Center for Artificial Intelligence (DFKI) and Saarland University, Saarland Informatics Campus, Saarbrücken, Germany.}%
\thanks{$^{3}$Tobias Dirndorfer, Alexander Kammenhuber, and Christoph Lauer are with the AUDI AG, Development Automated Driving, Ingolstadt, Germany.}%
}

\begin{document}

\bstctlcite{IEEEexample:BSTcontrol}
\maketitle
\thispagestyle{empty}
\pagestyle{empty}

\ifieeenotice
\CopyrightNotice{}
\fi

\tikzstyle{box}  = [shade,shading=axis,rectangle,draw,node distance=4mm and 4mm]%
\tikzstyle{boxb} = [box,rounded corners,top color=Cerulean,bottom color=MidnightBlue]%
\tikzstyle{boxg} = [box,rounded corners,top color=cyan,bottom color=cyan]%
\newcommand{\architecture}{%
    \node[boxg] (box0) {input};%
    \node[boxg] (box1) [right=of box0, align=center]{street data\\processing};%
    \node[boxg] (box2) [right=of box1, align=center]{trajectory\\generation};%
    \node[boxg] (box3) [right=of box2, align=center]{collision\\recognition};%
    \node[boxg] (box4) [right=of box3, align=center]{risk\\assessment};%
    \node[boxg] (box5) [right=of box4, align=center]{output};%
    \path[very thick,->] (box0) edge (box1);%
    \path[very thick,->] (box1) edge (box2);%
    \path[very thick,->] (box2) edge (box3);%
    \path[very thick,->] (box3) edge (box4);%
    \path[very thick,->] (box4) edge (box5);%
    \node[fit=(box1) (box4)] (box-cover) {};%
    \node[node distance=0mm and 0mm] (box-label) [above=of box-cover] {algorithm framework};%
    \begin{scope}[on background layer]
        \node[boxb, fit=(box-cover) (box-label)] (box-background) {};%
    \end{scope}
}%

\begin{abstract}
Due to the current developments towards autonomous driving and vehicle active safety, there is an increasing necessity for algorithms that are able to perform complex criticality predictions in real-time.
Being able to process multi-object traffic scenarios aids the implementation of a variety of automotive applications such as driver assistance systems for collision prevention and mitigation as well as fall-back systems for autonomous vehicles.

We present a fully model-based algorithm with a parallelizable architecture. The proposed algorithm can evaluate the criticality of complex, multi-modal (vehicles and pedestrians) traffic scenarios by simulating millions of trajectory combinations and detecting collisions between objects. The algorithm is able to estimate upcoming criticality at very early stages, demonstrating its potential for vehicle safety-systems and autonomous driving applications. An implementation on an embedded system in a test vehicle proves in a prototypical manner the compatibility of the algorithm with the hardware possibilities of modern cars. For a complex traffic scenario with 11 dynamic objects, more than $86$ million pose combinations are evaluated in $21$\,ms on the GPU of a Drive PX~2.
\end{abstract}


\section{Introduction and motivation}\label{sec:intro}

The number of vehicles on the road is constantly increasing.
According to Sousanis~\cite{sousanis2011world}, the billion units mark was passed in 2010, and, with the current growth rate, the tendency appears to remain the same for years to come.
With an increasing number of vehicles on the road, there is a rising interest and necessity of analyzing and planning trajectories in complex multi-modal traffic scenarios.
One goal for trajectory planning is the implementation of vehicle active safety systems for collision avoidance and mitigation~\cite{muller2016statistical}.
Globally, there are more than 1.2 million traffic-accident related fatalities per year~\cite{world2015global}.
Another goal is to enable fall-back or self-supervising systems for autonomous vehicles.

In this paper, we define \enquote{complexity} as the combination of the following factors: an undetermined number of moving objects to be considered, a large and undefined number of available options or \enquote{trajectories} these objects can follow, and an extensive variety of roads where vehicles can drive on.
Because of this complexity, the amount of information that has to be acquired, processed, and assessed for autonomous driving and vehicle active safety systems is accordingly huge.

One possible solution are machine learning methods.
A major advantage of these methods is the ability to analyze huge amounts of data in short time periods.
However, they also pose new challenges:
First, a comprehensive dataset is required in order to train machine learning models.
The compilation of a diverse and complete enough database, that ensures that each and every open street scenario is covered, results in an impractical approach due to the time and resources this would cost.
Current work in this area focuses only on a limited set of scenarios~\cite{bojarski2016end}.
Second, data labeling can be problematic if a supervised machine learning method is chosen.
Automated processes might not have the required classification accuracy for automotive safety systems and manual processes are very costly and time consuming.
Third, most machine learning models are not interpretable.
Rapidly evolving applications like cyber security profit greatly from methods like deep learning~\cite{McAfee}.
Yet, ethical and legal implications of false-positive actuations make machine learning methods inadequate for safety-critical applications.
Research on algorithm validation for automated driving functions focuses on a limited set of scenarios, too~\cite{8519598}.

Considering the previous points, we favor a model-based approach.
Here, the biggest challenge is the processing of complex and costly computations, which results in long execution times.
However, we will show that we can obtain execution times comparable to machine learning methods, but with deterministic outputs.
Since all objects (vehicles and pedestrians) are mobile, no information from previous simulations can be reused: each traffic scenario has to be simulated every time from ground up.
Even when the objects can be tracked over time, the trajectories that they are able to follow do change over time, altering the final outcome of the current traffic situation.
Considering this and the runtime constraints for applications for vehicle safety systems, it is imperative to design the algorithm for parallel processing.

In this paper, we introduce an approach that deals with the task of traffic analysis for criticality estimation and motion planning in a fully model-based fashion.
As a result, a large number of possible trajectories for all traffic participants have to be computed and assessed.
For this work, we define \enquote{criticality} as the probability that the current traffic situation will lead to a collision for the own (EGO-) vehicle.
The huge advantage of such an approach is that it can be easily tested, modified, and validated in simulation, which are key components for vehicle active safety algorithms.
The challenge of this approach is the high computational costs related to predicting trajectories in multi-object scenarios.
However, we will show that such an approach can run---when executed in parallel---in $21$\,ms on embedded hardware that is available for modern cars.

This paper makes the following contributions:

\begin{inparaenum}[1)]
    \item We present an algorithm for criticality estimation of complex, multi-modal traffic scenarios, that is novel for combining (i) vehicle models, steering controllers, acceleration profiles, and mechanical latencies for realistic trajectory generation (\autoref{sec:TGeneration}), (ii) the possibility of accurate modeling the geometry of objects for an accurate collision recognition (\autoref{sec:ColRec}), (iii) valid approximations for the conditional probabilities of collisions (\autoref{sec:Risk}) (iv) stochastic algorithm outputs and (v) all this while maintaining a highly parallelizable algorithm architecture.
    \item We present an efficient mapping of the proposed algorithm to embedded systems utilizing data-parallel GPU processing and unified system memory (\autoref{sec:mapping}).
    \item We show that our algorithm is able to simulate over $2000$\,trajectories, to recognize collisions between objects and to estimate the criticality of traffic scenarios in $21$\,ms (\autoref{sec:results}). This requires to evaluate more than $86$\,million pose combinations of the EGO-vehicle and the other mobile objects (Collision Objects - COs).
\end{inparaenum}

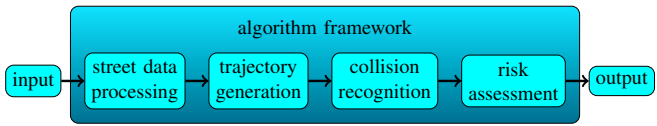
\begin{figure}
    \vspace{2 mm}
    \centering
    \resizebox{\columnwidth}{!}{%
        \begin{tikzpicture}%
        \architecture%
        \end{tikzpicture}%
    }%
    \caption{Architecture of the algorithm and sub-process division.}
    \label{fig:algo:architecture}
\end{figure}

\section{Related Work}\label{sec:RelWork}

Trajectory planning and risk assessment are topics of continuous research in the automotive area. These overlap, but differ on their main task. The first one focuses on finding a route A to B that fulfills certain criteria that can include collision avoidance, but tends to neglect certain possible trajectories for sake of efficiency. The second one focuses on estimating the criticality of a traffic situation, for which the future states of the objects are \textit{somehow} planned and/or predicted.

The work at hand is based on existing basic ideas like trajectory generation and collision recognition. It addresses key critical points of existing approaches that a) have been avoided by other authors because of the complexity or computational cost, b) were not considered combined until now and c) impact noticeably the final criticality estimation. The algorithm also differentiates itself by its flexible, highly parallelizable architecture, which is a base requirement for GPU-parallel mapping. These critical points are addressed following by comparing the present work with related publications.

Broadhurst et al.~\cite{broadhurst2005monte} present a method for \textit{reasoning} how vehicles move, when in presence of other traffic participants. The authors neglect several relevant aspects of the trajectory generation, such as mechanical latencies and limits. A purely geometrical model is used for describing vehicle motion, which generates non-driveable trajectories. This and the use of constant steering instead of a steering controller shift greatly the scenario criticality as shown in \autoref{fig:results}. The use of random components also complicates the algorithm validation for use in vehicle safety functions.

Broadhurst et al.~\cite{broadhurst2005monte} is a very adequate example of why it is necessary to approximate conditional probabilities as we will shown in \autoref{subsec:conditional}. As stated by Broadhurst et al.\ in Section II-F, the collision probability is calculated sequentially, which is time consuming. Their algorithm applied to a single object operates at $2$ Hz, while the algorithm presented in this work processes $30,869$ trajectory combinations of 4 objects (EGO + 3 CO) in the same time period.

Lef{\`e}vre et al.~\cite{lefevre2014survey} present a comprehensive survey on existing criticality estimation methods. They perform semantic classification and show trade-offs of the different classes. The results of this paper indicate that by using a smart, parallelizable algorithm structure, it is possible to obtain criticality estimations with a quality similar to that of \textit{interaction-aware} motion models (\autoref{sec:Risk}), but using \textit{maneuver-based} models (\autoref{sec:TGeneration}), which opens the door for real-time risk assessment.

Ziegler et al.~\cite{ziegler2014trajectory} and Ferguson et al.~\cite{ferguson2008motion} present real-world tested motion planning algorithms. Both rely on representing the motion planning as a constrained optimization task, where the trajectories are represented by cost functions that include a \enquote{no collision} constraint. The trajectory given by the optimized cost function is driven, even if it is the \textit{only} collision-free trajectory. This implies that errors in the sensor information or actuator control, unexpected CO behavior, modeling inaccuracies or wrong assumptions could lead to collisions. So, in highly dense traffic situations, unnecessary risks might be taken since only a small subset of the physically feasible trajectories are considered. That a trajectory is \textit{unlikely} to happen, does not mean that it \textit{cannot} happen.

Note that the goal of the present algorithm is \textit{not} to completely substitute current path planning or collision recognition approaches. Redundancy is a best practice where humans can be harmed due to system malfunction. As stated in \autoref{sec:intro}, the present algorithm can serve as a fall-back, supervising or plausibility system for both autonomous and human-driven vehicles.

\section{Multi-Hypotheses Approach}\label{sec:complexsection}

For estimating the criticality of a traffic scenario, an algorithm that can model interactions between multiple objects present in an area of interest is necessary. Not knowing exactly which trajectory will be followed by each of these objects generates uncertainty about the future. However, this uncertainty can be modeling by generating multiple hypotheses for each object. This addresses the possible motion options of the objects. Considering the previous, this algorithm was designed as a fully model-based multi-hypothesis algorithm.

A model-based approach has key benefits for passive and active vehicle safety systems. First, it enables the future integration of infrastructural elements (such as more lanes) and a larger number of static and dynamic objects in the computation. This could improve the precision, reliability and relevance of the obtained results. Second, this approach is based on domain knowledge and widely accepted mathematical models. This ensures deterministic outputs, allowing an easy validation of the algorithm. Finally, the model-based generation of statistical quantities for prediction tasks is important for automotive safety systems.

We define a \enquote{hypothesis} as the combination of a specific acceleration profile and a path that correlate over time; that is, one unique trajectory. A main advantage of the multi-hypothesis strategy is that the number of threads that are executed in parallel is easy to influence, as it depends on easily adjustable parameters such as the number of hypotheses
to be simulated or the amount of objects to be considered.
With this in mind, it is important to note that the more hypotheses that are simulated, the better coverage of all
possible states that an object might traverse in the near future. Therefore, the maximum possible accuracy of the algorithm can be scaled depending on available parallel computing resources.

The algorithm is composed by four modules which have to be executed in a given order, and
inside these modules, there are tasks that can be executed in parallel. The
modules and the computational structure of the algorithm can be
seen in \autoref{fig:algo:architecture}.

To manage the parallel execution of the algorithm, a set of matrices were structured to keep the information traceable and independent, and nested
indexes had to be generated to allow the information to be
retrieved. These are module-specific and it is precisely these
indexes that express the parallelization possibilities of the
corresponding tasks. This will be explained in the corresponding modules.

\section{Algorithm Process}\label{sec:algorithm}

In order to detect collisions between two objects, their pose (coordinates of center of gravity and orientation) over time (trajectory) and their shape has to be known. This means that each and every prediction has to go through all four modules, which will be explained in the following.

\subsection{Street Data Processing}\label{sec:StreetData}

One of the constraints for the trajectory generation is the road
infrastructure, meaning that the vehicles are bound to it under conventional traffic circumstances (e.\,g. not
going off-road). Because of this, the trajectories will be generated
according to the lanes present on the driveable road. Considering
that the focus of this work is not the street modeling, the
assumption is made that the lane information is delivered
from the sensors to the algorithm in the form of
sets of three pairs of (x,\,y) coordinates. Using exteroceptive sensors in
vehicles, this representation is realized in modern
cars. Each one of these sets corresponds to a specific lane divider, and each lane is delimited by exactly two lane dividers. Adjacent
lanes share one lane divider.

\begin{figure}
\vspace{1 mm}
    \centering
    \includegraphics[width=\columnwidth]{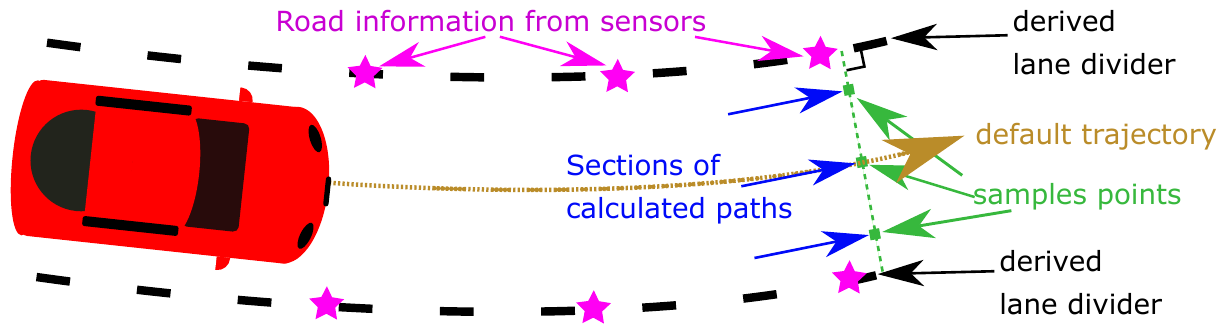}
    \caption{Graphical depiction of the trajectory generation and lane divider derivation.}
    \label{fig:street_modeling}
\end{figure}

From each set, it is assumed that two coordinates correspond to
the closest and farthest detected points of the lane divider, and
the third one is any point in between. From these three
points and using the Gauss-Jordan method~\cite{gaussjordan}, a second degree
equation is obtained, which corresponds to the mathematical
representation of the associated lane divider.
A graphical explanation of this can be seen in \autoref{fig:street_modeling}.

The algorithm is designed to
consider up to three lanes: the own (EGO-), immediate left, and immediate right
lanes. If the sensors do not deliver information for neighbouring lanes, they will not be considered, and no trajectory will be generated in the corresponding area. If the information of the EGO-lane is missing, a \enquote{virtual} lane will be generated according to the current vehicle state and applicable legislation~\cite{RAADeu}.

\begin{figure}
\vspace{2 mm}
    \centering
    \includegraphics[width=\columnwidth]{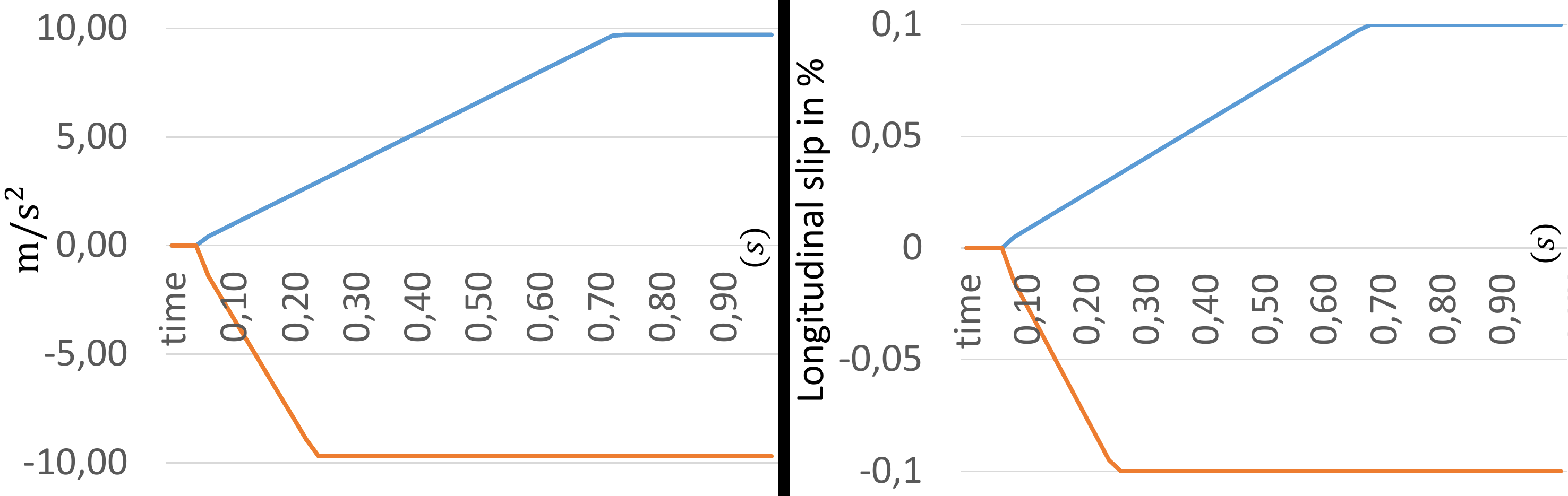}
    \caption{Examples of acceleration profiles for the one-track (OT) model (left) and longitudinal slip profiles for the two-track (TT) model (right).}
    \label{fig:AccelProfiles}
\end{figure}

Having the mathematical representation of the driveable road, the vehicles are associated with their corresponding lanes. The EGO-vehicle will always be placed on the EGO-lane. All other vehicles are associated with the lanes according to the location
of their estimated center of gravity (more on vehicle parameters later).
Pedestrians are not bound to the road infrastructure, so no association is performed for them.

For this module, up to 4 threads can be computed in parallel, one for each lane divider.

\subsection{Trajectory Generation}\label{sec:TGeneration}
Having the mathematical representation of the road (\autoref{sec:StreetData}), many hypotheses for each object are computed by means of motion models as described in the following.
\subsubsection{Motion Models for Vehicles} \label{sec:ModelVehicles}

Under tractive driving (i.\,e. not
exceeding the grip limits of the tires), longitudinal and lateral dynamics of the vehicles are coupled. For this, motion models are used for generating the trajectories of the vehicles.

\begin{inparaenum}[a)]

\item The trajectories of the EGO-vehicle are computed using the TT model~\cite{mitschke1972dynamik}, since all relevant information is known. Its mathematical representation is given by
\begin{equation*} \label{twotrack}
\begin{split}
\boldsymbol{\dot{x}}=
\begin{bmatrix}
\dot{v}&
\dot{\beta}&
\ddot{\psi}
\end{bmatrix}^T=
\boldsymbol{f}\left(v,\beta,\dot{\psi},\boldsymbol{F}_{\text{l}ij},\boldsymbol{F}_{\text{s}ij}\right),
\end{split}
\end{equation*}
where $\beta$ is the sideslip angle, $\psi$ is the yaw angle in global frame, $v$ is the current velocity over ground, $\boldsymbol{F}$ represents the tire forces, l and s indicate longitudinal and side forces, and indices $i$ and $j$ denote the front, rear, left, and right tire.

\item For all other vehicles, a OT model~\cite{schramm2014vehicle} is chosen since, up to now, it is not possible to acquire all the information required
for the TT model implementation in real-time.
The used OT model is given by
\begin{equation*} \label{onetrack}
\begin{split}
\begin{bmatrix}
\dot{\beta}\\
\ddot{\psi}
\end{bmatrix} \!=\! &\setstretch{1.25}
\begin{bmatrix}
-\frac{c_{\alpha f}+c_{\alpha r}}{mv} & \frac{c_{\alpha r}l_r-c_{\alpha f}l_f}{mv^2}-1\\
\frac{c_{\alpha r}l_r-c_{\alpha f}l_f}{I_z} & -\frac{c_{\alpha f}l_f^2+c_{\alpha r}l_r^2}{I_zv}
\end{bmatrix}
\begin{bmatrix}
\beta\\
\dot{\psi}
\end{bmatrix} \!+\!
\begin{bmatrix}
\frac{c_{\alpha f}}{mv}
\\
\frac{c_{\alpha f}l_f}{I_z}
\end{bmatrix}
\delta_f,
\end{split}
\end{equation*}
where $c_{\alpha f}$ and $c_{\alpha r}$ are the front and rear tire cornering stiffness, $l_f$ and $l_r$ indicate the distance from the center of gravity of the vehicle to the front and rear axles, $m$ and $I_z$ are the mass and moment of inertia of the vehicle, and $\delta_f$ is the front steering angle at the tires.
\end{inparaenum}

For both motion models, once having the corresponding state variables, the pose of the vehicle is calculated by means of Euler integration as described by the following equations:
\begin{align*}
\label{EulerVelocity}
\begin{split}
& \begin{bmatrix}
v_{V,n+1}\\
\beta_{n+1}
\end{bmatrix}=
\begin{bmatrix}
v_{V,n}\\
\beta_{n}
\end{bmatrix}+
\begin{bmatrix}
\dot{v}_{V,n}\\
\dot{\beta}_{n}
\end{bmatrix} \tau,\\
& \psi_{V,n+1}=\,
\psi_{V,n}+\dot{\psi}_{V,n} \tau+\ddot{\psi}_{V,n} \frac{\tau^2}{2},\\
& \begin{bmatrix}
a_{x,V,n+1}\\
a_{y,V,n+1}
\end{bmatrix} \!=\!
\begin{bmatrix}
c(\beta_{n})\\
s(\beta_{n})
\end{bmatrix} \! \dot{v}_{V,n}
\!+\! \begin{bmatrix}
-s(\beta_{n})\\
c(\beta_{n})
\end{bmatrix} v_{V,n} \left(\dot{\beta}_{n} \!+\!\dot{\psi}_{V,n}\right),\\
& \begin{bmatrix}
x_{V,n+1}\\
y_{V,n+1}
\end{bmatrix}=
\begin{bmatrix}
x_{V,n}\\
y_{V,n}
\end{bmatrix}
+
\begin{bmatrix}
c(\psi_{V,n}+\beta_{n})\\
s(\psi_{V,n}+\beta_{n})
\end{bmatrix} v_{V,n} \tau\\
& \hspace{1.4cm} +\! \begin{bmatrix}
c(\psi_{V,n})\\
s(\psi_{V,n})
\end{bmatrix} a_{x,V,n} \frac{\tau^2}{2} \!+\!
\begin{bmatrix}
-s(\psi_{V,n})\\
c(\psi_{V,n})
\end{bmatrix}
a_{y,V,n} \frac{\tau^2}{2},
\end{split}
\end{align*}
where $s$ and $c$ are the sine and cosine functions, subscript $V$ denotes a vehicle, $\tau$ is the time lapse between instances $n+1$ and $n$, $[x_{V},y_{V}]^T$ are the coordinates in global frame, $a_{x,V}$ and $a_{y,V}$ are the accelerations in vehicle coordinate frame. The input $\dot{v}_{V}$ will be addressed in \autoref{sec:VariationVehicleModel}.

The mass of the vehicles plays an important role in traffic accidents~\cite{iihs2009carsize} and it is a parameter for the OT model. Because of this, the COs are classified according to their dimensions to assign them a mass value. For this, it is assumed that the on-board sensors of the EGO-vehicle can deliver approximate information about the dimensions of the COs. If the height of the COs cannot be estimated, $6$ classes are used: quadricycle, supermini, small family car, large family car, executive, and multi-purpose vehicle. Otherwise, $2$ extra classes are used: off-roader and cargo. This accounts for the different length/width-to-weight ratios that taller vehicles have, when compared to the other classes. The classes are chosen according to Van Miert~\cite{cartype} and average values are obtained from Heydinger et al.~\cite{vehicleinertia}. Typical values for tire parameters and vehicle geometry are obtained from Isermann~\cite{isermann2006fahrdynamik}.

\subsubsection{Variation for the Vehicle Trajectory Generation} \label{sec:VariationVehicleModel}
The generation of multiple hypotheses requires to influence both longitudinal and lateral dynamics of the vehicles.
\begin{inparaenum}[a)]

\item For the longitudinal motion, either the proper acceleration or longitudinal slip is sampled. Two samples are the maximum and minimum of the corresponding range, and one is zero (vehicle is cruising). The remaining samples are equally distributed in the negative region to favor a reduction of kinetic energy of the bodies. This is very robust in critical situations. The positive area is also sampled to address
cases where accelerating would avoid an accident, like a
rear-end collision. The ranges go from $-9.7\frac{\si{\metre}}{\si{\second}^2}$ to $9.7\frac{\si{\metre}}{\si{\second}^2}$ (OT) and the longitudinal slip from $-0.1$ to $0.1$ (TT). Furthermore, mechanical latencies and jerks are introduced to generate profiles that are present in road vehicles. This aids to maintain realistic trajectories. The number of profiles are equal for the EGO-vehicle and COs. The output of the profiles is $\dot{v}_{V}$, which is an input for the corresponding motion model. Examples of these profiles can be seen in \autoref{fig:AccelProfiles}.

\item To get feasible and realistic trajectories, a controller is designed and implemented for lateral dynamics. First, a reference trajectory is generated with a motion model and the expected driver behavior. That is, the vehicles keep driving with their current acceleration towards the middle of their current lane. Along this trajectory, the available lanes are sampled at three predefined instances of the prediction time ($1.0 \si{\second}$, $1.5 \si{\second}$ and $2.0 \si{\second}$). The sample points are equally distributed on each lane and perpendicular to the lane divider. Three samples are made on the own lane and two samples are made on the neighboring lanes. This reflects the expected behavior of the vehicles: it is more likely that they will stay on their current lane, rather than steer towards the neighboring lanes. Out of these sample points, the path sections are generated. These sections go over the sample points and are parallel to the lane dividers (see \autoref{fig:street_modeling}). Three of these path sections (one from each sampling instance) represent one complete path. The path sections are the input for the controller and the output is the steering angle at the front wheels $\delta_f$. This is then used as input for the correspondent motion model. It was decided to generate more complete paths for the EGO-vehicle than for the COs. This takes into account that one is able to influence the EGO-vehicle, but not the COs.

The lateral controller has been optimized using a large number of simulations and is expressed mathematically by
\begin{equation*} \label{controller}
\begin{split}
\delta_f=&\left(\Bigl\lvert
\left(-0.018\cdot v_V\!+\!1.5\right)
d_{\text{path}}\Bigr\lvert+0.5\right)d_{\text{path}}\\&+\!\left(-\Bigl\lvert
\left(-0.018\cdot v_V\!+\!1.5\right)
d_{\text{path}}\Bigr\lvert+9.5\right)\left(3.8197\cdot e_{\psi}\right),
\end{split}
\end{equation*}
where $d_{\text{path}}$ is the
closest distance between the predicted center of gravity of the vehicle and the path section (distance error), and
$e_{\psi}$ is the difference between $\psi_V$ and the direction of the section of the calculated path (yaw error). So, the controller is realized based on a predicted error
and two P-controllers.
A graphical description of the functioning principle of the lateral controller can be seen can be seen in \autoref{fig:controller_idea}.

\begin{figure}
\vspace{2 mm}
    \centering
    \includegraphics[width=\columnwidth]{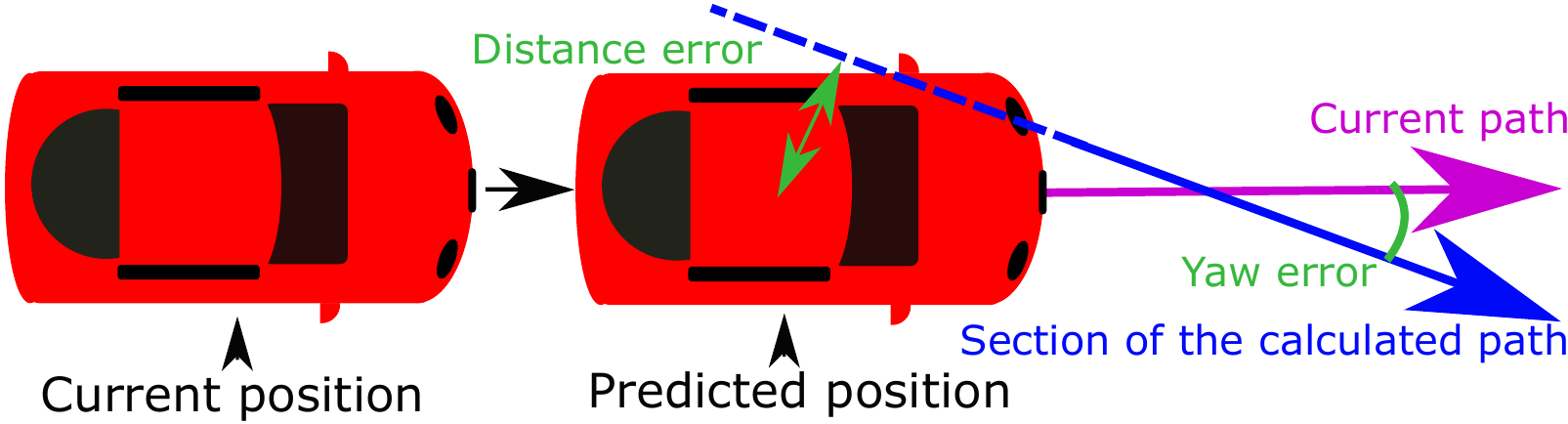}
    \caption{Principle of the lateral dynamics controller.}
    \label{fig:controller_idea}
\end{figure}

To optimize the performance of the controller, its design also includes features such as dynamically weighting $d_{\text{path}}$ and $e_{\psi}$, and dynamically adjusting the prediction time for the position of the vehicle according to $v_V$. A maximum steering angle and maximum steering rate are also included for considering the mechanical limitations of the road vehicles and driver capabilities~\cite{driversteer}.

\end{inparaenum}

\subsubsection{Motion Model for Pedestrians} \label{sec:ModelPedestrian}

Research on pedestrian motion modeling has been done by Schlake~\cite{pedestrian1}. Factors like emergency situations and pedestrian density do affect the motion of pedestrians, but these motion models focus on in-building situations. In open-air environments, the movement of pedestrians is not bound to the road infrastructure. Because of this, the pedestrian motion is expressed by a kinematic model with no controller:
\begin{equation*} \label{pedestrianmotion}
\begin{split}
\begin{bmatrix}
x_{P,n+1}\\
y_{P,n+1}
\end{bmatrix}=&
\begin{bmatrix}
x_{P,n}\\
y_{P,n}
\end{bmatrix}
+
\begin{bmatrix}
c(\psi_{P})\\
s(\psi_{P})
\end{bmatrix} v_{P,n} \tau+
\begin{bmatrix}
c(\psi_{P})\\
s(\psi_{P})
\end{bmatrix}
\dot{v}_{P,n} \frac{\tau^2}{2},
\end{split}
\end{equation*}
where subscript $P$ denotes a pedestrian. 

For pedestrians, the samples are equally distributed from $0\si{\degree}$ to $360\si{\degree}$ for $\psi_P$ and from $-12\frac{\si{\metre}}{\si{\second}^2}$ to $12\frac{\si{\metre}}{\si{\second}^2}$ for $\dot{v}_P$. The velocity $v_P$ is limited to $2.7\frac{\si{\metre}}{\si{\second}}$ assuming that pedestrian velocities on open roads range from slow to running, rather than sprinting~\cite{zkebala2012pedestrian}. The number of samples of $\psi_P$ is equal to the number of complete paths for COs and the number of samples of $\dot{v}_P$ is equal to the number of profiles for vehicles.

\subsubsection{Parallelization of Trajectory Generation} \label{sec:TrajPar}
A key for computing so much information in parallel is to generate the trajectories so that they are completely independent from each other. In \autoref{sec:TGeneration}, the generation of trajectories by combining acceleration profiles and paths is explained. It is precisely this combination that allows the generation of a large number of independent trajectories.

The total number of trajectories $r_{\text{tra}}$ that can be simulated in parallel is equal to
\begin{equation*} \label{threads:traj}
r_{\text{tra}}=\left(o\cdot h_{\text{acc}}\cdot h_{\text{CO,str}}\right)
+\left(h_{\text{acc}}\cdot h_{\text{EGO,str}}\right),
\end{equation*}
where $o$ is the total
number of COs considered,
$h_{\text{acc}}$ is the number of acceleration
profiles, $h_{\text{CO,str}}$ is the number of
complete paths for the COs, and
$h_{\text{EGO,str}}$ is the number of complete paths for the EGO-vehicle.

In this paper, the number of trajectories for the EGO-vehicle is equal to $2058$. This results from the linear combination of the path sections $(h_{\text{EGO,str}}=7^3)$ which are also combined with $6$ acceleration profiles. As stated in \autoref{sec:VariationVehicleModel}, one is not able to influence the COs. For this reason, no combination of the path sections is made for COs. This means that each CO has $7$ complete paths, which are combined with $6$ acceleration profiles to generate $42$ trajectories. For example, $2478$ trajectory generations can be executed in parallel when $o=10$.

It should be noted that the used motion models are represented by differential equations that are solved numerically for each time instance of each trajectory (\autoref{sec:ModelVehicles}). This means that the current position has to be known in order to calculate the future position of an object. Following this train of thought, one limitation of parallelizing is that it is not possible to simulate all time instances of one single trajectory in parallel---they have to be calculated sequentially.

\subsection{Collision Recognition}\label{sec:ColRec}

\subsubsection{Object Modelling with Polygons}

Once the trajectories are generated, it is possible to check if a collision between the EGO-vehicle and a CO occurs.
For this, the complete set of hypotheses of each CO is combined with each and every hypothesis of the EGO-vehicle.
Both objects are then modeled as polygons, and it is checked at each time instance whether they overlap or not.
If the polygons overlap, this indicates that a collision occurs at the given time instance.
Well known point in polygon strategies exist to test for this~\cite{shimrat1962algorithm,haines1994point}.
The method requires to check the overlapping twice per time instance: EGO over CO, and CO over EGO.
The polygons can be as complex as necessary. The more complex they are, the better the objects
are described, but the more runtime is required. This is
very relevant for the computational resources, since the
collision recognition is the module that takes the most runtime and the overlapping check is the function that is called most
frequently.

Combinations of hypotheses between COs are not
taken into account, seeing that it is not the focus of this work to
predict collisions that do not involve the EGO-vehicle.

\subsubsection{Parallelization of Collision Recognition}

As mentioned in \autoref{sec:TrajPar}, all trajectories are completely independent from each other. Thus, any combination resulting from them is independent as well to be simulated and evaluated by different processing units in parallel. The resulting number of trajectory combinations $r_{\text{col}}$ that could be simulated in parallel is given by
\begin{equation*} \label{threads:coll}
\begin{split}
r_{\text{col}}=o\cdot h_{\text{CO,str}}\cdot h_{\text{EGO,str}}\cdot h_{\text{acc}}^2.
\end{split}
\end{equation*}
As an example, for a case with $o=10$, $h_{\text{acc}}=6$, $h_{\text{EGO,str}}=343$ and $h_{\text{CO,str}}=7$, $864,360$ trajectory combinations are executed in parallel in this module in $21$\,ms.

In this work, a maximum prediction time of $2$\,s and a discretization time of $20$\,ms is used for any given scenario. This means that the overlapping check function will be called 200 times for each trajectory combination, yielding a total number of $172,872,000$ calls per scenario for this routine alone that are executed in $21$\,ms for this example.

The length and width used for cars in the class with the smallest vehicles is $3.13$\,m and $1.46$\,m, respectively. Assuming the two specific collision cases of a T-bone and a purely longitudinal one, the vehicles need a relative velocity of $229\frac{\si{\metre}}{\si{\second}}$ and $313\frac{\si{\metre}}{\si{\second}}$ accordingly for them to drive "through" each other and the collision to be undetected. It is known to the authors that the collision configurations are infinite, and that special cases, such as a small overlap, do happen. However, this shows that the discretization time covers a very comprehensive range of street scenarios.

\subsection{Risk Assessment}\label{sec:Risk}

The risk assessment is based on the anticipation time and criticality. We define \enquote{anticipation time} as the time in advance that the criticality of a situation can be recognized. 
For this, once it is determined that a combination of trajectories
leads to a collision, the probabilities of these trajectories are multiplied, obtaining the probability that this combination will occur. Assuming statistical independence (\autoref{subsec:conditional}), the probability $p_{\text{cra}}$ that
a traffic scenario will lead to a collision is given by
\begin{equation*} \label{totalCrashProb}
\begin{split}
p_{\text{cra}}=&\sum_{i=1}^{r_{\text{EGO}}}
\sum_{j=1}^{r_{\text{CO}}}
I(i,j)\cdot p_{\text{EGO,}i}\cdot p_{\text{CO,}j},
\end{split}
\end{equation*}
where $I(i,j)$ is the indicator function that yields
one if and only if the $i$-th EGO and the
$j$-th CO trajectory lead to a collision and zero
otherwise,
$r_{\text{EGO}}=h_{\text{acc}}\cdot h_{\text{EGO,str}}$,
$r_{\text{CO}}=o\cdot h_{\text{acc}}\cdot h_{\text{CO,str}}$,
$p_{\text{EGO,}i}$ is the probability of the
$i$-th EGO trajectory, and
$p_{\text{CO,}j}$ is the probability of the
$j$-th CO trajectory.

\subsubsection{Trajectory Probability Calculation}\label{subsec:traj:prob:calc}

To get the probability that a specific hypothesis can occur, it is scored against the reference trajectory mentioned in \autoref{sec:VariationVehicleModel}. The lower the score, the lower the occurrence probability.
The scoring value $n_{\text{h}}$ is given as follows:
\begin{equation*} \label{normTraj}
\begin{split}
n_{\text{h}}=&\cfrac{\left[w_{\text{acc}}\cdot n_{\text{acc}}\right]+\left[w_{\text{str}}\cdot d_{\text{str}}\right]}{c_{\text{com}}\cdot c_{\text{cou}}},
\end{split}
\end{equation*}
where $c_{\text{com}}$ and
$c_{\text{cou}}$ are penalizing factors for the
complexity of the maneuver and for entering lanes with counter
traffic; $n_{\text{acc}}$ and
$d_{\text{str}}$ are the scoring factors for
the acceleration profile and the path, and
$w_{\text{acc}}$ and
$w_{\text{str}}$ are fixed weighting factors for the
acceleration and steering correspondingly. The scoring and penalizing factors steer the occurrence probability of the hypotheses, thus the criticality of the traffic scenario as well. Should colliding trajectories have a higher occurrence probability, the criticality of the scenario will be higher too. One way of optimizing the parameters in practice, is according to the expected passenger injury for each type of collision. The used parameters are chosen using domain knowledge.

Once all trajectories of an object are scored, the scores are normalized with respect to the $L^1$-norm. The resulting values are considered as occurrence probabilities of the trajectories. This process is repeated for each object.

\subsubsection{Conditional Probability}\label{subsec:conditional}

In \autoref{subsec:traj:prob:calc}, an occurrence probability for each
hypothesis of each object is calculated. Having multiple COs that can collide with EGO at different time instances, the conditional probability of these collisions has to be considered. The implementation of
conditional probabilities could give a marginal benefit for representing the
behavior of the criticality, but the required mathematics prevent the
algorithm from being parallelizable.

To maintain the algorithm parallelizable, the conditional probability is approximated. For this, it is assumed that an EGO-CO hypothesis combination can occur if and only if no other collision occurred before along the corresponding EGO-trajectory. Thus, the EGO-CO combinations are considered to be independent, and their probabilities are scaled down in chronological order. That is, the first hypothesis combination to occur in time maintains its estimated probabilities (\enquote{collision} and \enquote{no collision}). Then, the probability that the next hypothesis combination occurs, is equal to the \enquote{no collision} probability of the first combination. Both \enquote{collision} and \enquote{no collision} probabilities are then scaled down accordingly. This process continues until all the probabilities of all the combinations are scaled.

\begin{figure}
\vspace{2 mm}
    \centering
    \includegraphics[width=0.8\columnwidth]{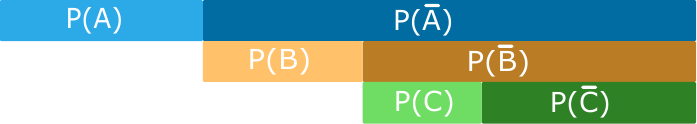}
    \caption{Graphical exemplification of the implemented approximation of conditional probability.}
    \label{fig:cond_prob}
\end{figure}

The following equations explain this further and a graphical representation of this can be seen in \autoref{fig:cond_prob}
\begin{equation*} \label{p21}
\begin{split}
&P\left(B\cap \bar{A}\right)\approx P\left(
\bar{A}\right)\cdot P\left(B\right)\\
&P\left(C\cap \bar{B} \cap \bar{A}\right)\approx
P\left(\bar{A}\right)\cdot P\left(\bar{B}\right) \cdot P\left(C\right).
\end{split}
\end{equation*}

\section{Algorithm Mapping to Embedded Systems}\label{sec:mapping}

The proposed algorithm is designed such that it can be executed in parallel by thousands of threads.
While each module in \autoref{fig:algo:architecture} depends on the results of its predecessor, the modules themselves are highly parallelizable.
We target embedded systems that offer embedded, on-board GPUs as execution platform.
For automotive applications, there is a variety of hardware platforms offered by different manufacturers such as Renesas (R-Car), NXP (i.MX), Texas Instruments (OMAP Jacinto), Qualcomm (Snapdragon), Intel (GO), or NVIDIA (Jetson).

For real-time performance, it is essential to map the algorithm efficiently to the available hardware resources and to make best use of them.
This requires typically low-level programming and ties the implementation to one specific architecture.
To avoid this, we use the AnyDSL\footnote{\url{https://anydsl.github.io}} framework that allows to separate low-level hardware-specific aspects from the high-level algorithm description~\cite{leissa2018anydsl,leissa2015shallow}.
AnyDSL supports code generation for GPUs by generating CUDA and OpenCL.
The algorithm description for CPU and GPU are the same, only a hardware-specific mapping is required for each target platform:
\begin{itemize}
    \item Iteration Logic: Defines in which order data is processed in each module. This can be sequential on the CPU vs.\ data-parallel on the GPU.
    \item Hardware Intrinsics: Many trigonometrical functions such as sine or cosine can be mapped to much faster hardware-accelerated versions on the GPU.
    \item Memory Hierarchy: Modern CPUs and GPUs have a deep memory hierarchy. Some of them require explicit programming.
    \item Memory Management: CPU and GPU share the same physical memory in most embedded systems.
\end{itemize}

In our implementation, we exploit all those hardware-specific features mapping all modules to the GPU.
In particular, the resulting implementation executes all modules in a data-parallel fashion, makes extensive use of hardware intrinsics for the collision recognition (sine, cosine, tangent), and requires no data-transfers between CPU and GPU, exploiting unified CPU/GPU memory.
For deployment, we use Thrift\footnote{\url{https://thrift.apache.org}} to retrieve input data from the vehicle and return the results of to our algorithm.

\section{Evaluation and Results}\label{sec:results}

\subsection{Algorithm Outputs}\label{sec:algo:outputs}

\begin{figure}
\vspace{2 mm}
    \centering
    \includegraphics[width=\columnwidth]{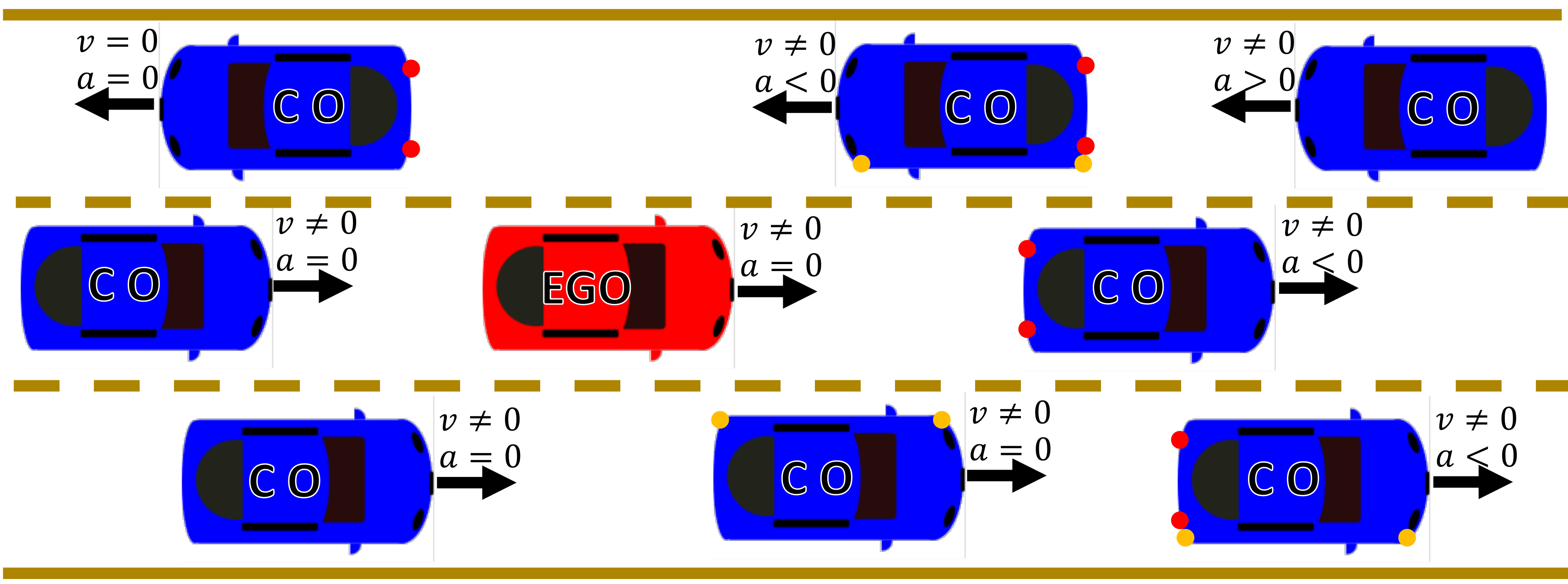}
    \caption{Example of a complex traffic scenario that can be handled by the presented algorithm.}
    \label{fig:scenarioexm}
\end{figure}

The evaluation is performed first by designing a set of 20 different simulated scenarios that cover in a wide, general manner possible traffic situations. These scenarios include combinations of one, two, and three lanes, COs as static and moving objects, as well as counter traffic and vehicles approaching from behind. The Figure \autoref{fig:scenarioexm} shows an example of a complex traffic scenario that can be handled by the proposed algorithm. The second part of the evaluation of the algorithm includes 547 real-life traffic situations. For this, a vehicle is driven on open roads and the information obtained from the sensors is collected and used as an input for the algorithm for off-line evaluation.

An evaluation metric of the algorithm stability is the development of the predicted criticality over time. A smooth, progressive development indicates the absence of misjudged collisions (false-positives). It includes in a compact and understandable manner the most relevant information about the traffic scenario for passive and active vehicle safety systems. This because features like false-positives and triggering thresholds are recognized easily.

The obtained results indicate that the algorithm is capable of detecting unavoidable collisions with an anticipation time that depends on the scenario that is being evaluated. When the traffic is purely longitudinal, this anticipation time is long enough to influence the vehicle dynamics. When cross-traffic is present, this anticipation time is long enough to activate passive safety systems.
For the designed scenarios where static objects were placed in different arrangements in front of the EGO-vehicle, the average anticipation time was $932$\,ms; and for the scenarios where the COs were in motion, it ranged from $660$\,ms to $1800$\,ms. These results contrast with the $300$\,ms that occur in a simple scenario without road modeling. The large anticipation times could aid a better triggering of multistage airbags, thus preventing passenger injuries~\cite{iihs2004evidence}. This is specially interesting for lateral airbags~\cite{iihs2003headprotecting}. Slow actuators could benefit from this too, since the extra time helps to compensate mechanical latencies.

\begin{figure}
\vspace{2 mm}
    \centering
    \includegraphics[width=\columnwidth]{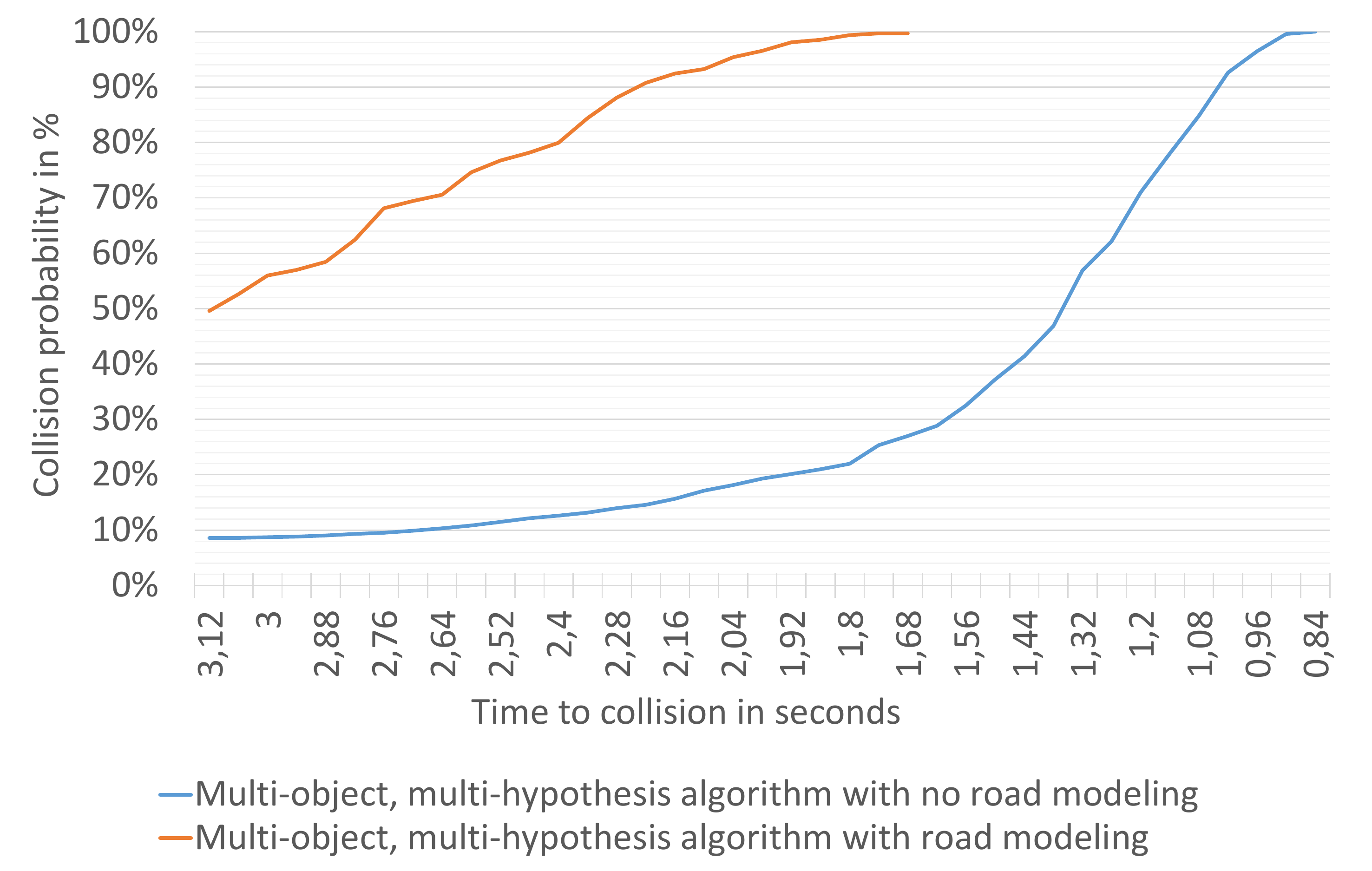}
    \caption{Development over time of the predicted collision probability of a single test scenario when evaluated by two algorithms.}
    \label{fig:results}
\end{figure}

One of the pillars of our algorithm is the combination of road modeling, vehicle motion models and the lateral dynamics controller. This ensures that the generated trajectories are not only drivable by vehicles, but that they are meaningful as well. \autoref{fig:results} shows a good comparison of this. For evaluation purposes, the controller module is disabled. This provokes that the trajectory generation does not take into consideration the drivable road, and that some of the trajectories tagged as \enquote{collision-free} actually go outside the available road infrastructure.

An extremely important result is that there are zero false-positive outputs for the evaluated scenarios. This demonstrates that the algorithm possesses a high degree of reliability and robustness. This is specially relevant when deciding triggering points of passive and active vehicle safety systems.

An additional benefit of the algorithm is the output of possible escape routes, that is, trajectories that could aid to avoid an oncoming collision. This derives from two factors. First, as stated in \autoref{sec:TGeneration}, all the trajectories are generated by the use of motion models combined with the aid of realistic acceleration profiles and a controller for lateral dynamics. This means that all generated and simulated trajectories can be driven by a vehicle. Second, as described in \autoref{sec:Risk}, the complete probability spectrum of the EGO-trajectories is known, so the most adequate one can be chosen. As stated in \autoref{sec:complexsection}, it is important to note that the resolution of the algorithm depends highly on the amount of acceleration profiles and paths (hypotheses).

\subsection{Execution Time}

For evaluation, we use the Drive PX~2 development board from NVIDIA.
The Drive PX~2 has a CPU with four ARM Cortex A57 cores and two NVIDIA Denver cores as well as a Tegra X2 GPU (GP10B) with 256 cores and a dedicated GPU (GP106) with 1152 cores, both based on the Pascal architecture.
While the Tegra X2 GPU shares the main memory of the CPU, the dedicated GPU has its own memory.

We consider two scenarios: The first scenario (S1) considers 3 COs in addition to the EGO-vehicle while the second scenario (S2) considers 10 COs in addition to the EGO-vehicle.
We use $6$ acceleration profiles, which are combined with $7$ and $343$ paths for CO and EGO-vehicle, respectively.
For each scenario, we compute the trajectories for the next $2$ seconds with a resolution of $20$ ms, which equals to $100$ time steps.
This results in $8.64$ million pose combinations that need to be evaluated per CO.
In total, this results in $25.93$ million pose combinations for S1 and $86.44$ million pose combinations for S2.
Using only $5$ acceleration profiles reduces the number of pose combinations to $18.00$ million for S1 and $60.01$ million for S2, respectively.
The execution time for scenario S1 and S2 is shown in \autoref{tab:results}.
On the dedicated (GP106) / embedded (GP10B) GPU, it takes $10$/$15$\,ms to evaluate the proposed algorithm for S1 and $21$/$49$\,ms for S2.
Considering only $5$ acceleration profiles reduces the execution time to $8$ /$11$\,ms for S1 and $14$/$35$\,ms for S2.
More than two thirds of the total execution time is spent for collision recognition.
The GPU execution is more than two magnitudes faster than CPU execution, benefiting from hardware-accelerated trigonometrical functions.
The communication overhead of the client/server architecture of Thrift adds $0.5$\,ms on top of the algorithm execution time.

\newcommand{\mc}[1]{\multirow{2}{*}{#1}}
\ctable[mincapwidth=\columnwidth,pos=h,label=tab:results,doinside=\relscale{0.94},
caption={Performance results for the proposed algorithm. Shown is the number of pose combinations evaluated as well as the median execution time in \textbf{ms} (lower is better) on the CPU, dedicated GPU (GP106), and embedded GPU (GP10B).}]{lrrrr}{}{
                           \FL
Scenario        &       \# pose &      CPU  &   GPU & GPU   \NN
                &  combinations &           & GP106 & GP10B \ML 
\mc{S1 (3 CO)}  & 25.93 million &  1800 ms  & 10 ms & 15 ms \NN 
                & 18.00 million &  1600 ms  &  8 ms & 11 ms \ML 
\mc{S2 (10 CO)} & 86.44 million & 11600 ms  & 21 ms & 49 ms \NN 
                & 60.01 million &  9600 ms  & 14 ms & 35 ms \LL 
}

\section{Conclusions}

In this work, a fully model-based multi-modal parallelizable algorithm is presented. This algorithm is able to estimate upcoming criticality of complex traffic scenarios at very early stages. The architecture of the algorithm allows the further inclusion of road infrastructure and mobile objects. This architecture also allows the algorithm to be ported to different GPUs. The implementation on vehicle-compatible hardware proves in a prototypical manner its feasibility to function in production vehicles. Short execution times, deterministic results, and the absence of false-positives, prove the adequacy of the algorithm for passive and active vehicle safety systems.

\bibliographystyle{IEEEtran}
\bibliography{literature}

\begin{thebibliography}{10}
\providecommand{\url}[1]{#1}
\csname url@rmstyle\endcsname
\providecommand{\newblock}{\relax}
\providecommand{\bibinfo}[2]{#2}
\providecommand\BIBentrySTDinterwordspacing{\spaceskip=0pt\relax}
\providecommand\BIBentryALTinterwordstretchfactor{4}
\providecommand\BIBentryALTinterwordspacing{\spaceskip=\fontdimen2\font plus
\BIBentryALTinterwordstretchfactor\fontdimen3\font minus
  \fontdimen4\font\relax}
\providecommand\BIBforeignlanguage[2]{{%
\expandafter\ifx\csname l@#1\endcsname\relax
\typeout{** WARNING: IEEEtran.bst: No hyphenation pattern has been}%
\typeout{** loaded for the language `#1'. Using the pattern for}%
\typeout{** the default language instead.}%
\else
\language=\csname l@#1\endcsname
\fi
#2}}

\bibitem{sousanis2011world}
J.~Sousanis, ``World vehicle population tops 1 billion units,'' \emph{Wards
  Auto}, vol.~15, 2011.

\bibitem{muller2016statistical}
M.~M{\"u}ller, P.~Nadarajan, M.~Botsch, W.~Utschick, D.~B{\"o}hml{\"a}nder, and
  S.~Katzenbogen, ``A statistical learning approach for estimating the
  reliability of crash severity predictions,'' in \emph{19th International
  Conference on Intelligent Transportation Systems (ITSC)}.\hskip 1em plus
  0.5em minus 0.4em\relax IEEE, 2016, pp. 2199--2206.

\bibitem{world2015global}
{World Health Organization}, \emph{Global Status Report on Road Safety
  2015}.\hskip 1em plus 0.5em minus 0.4em\relax World Health Organization,
  2015.

\bibitem{bojarski2016end}
M.~Bojarski, D.~Del~Testa, D.~Dworakowski, B.~Firner, B.~Flepp, P.~Goyal, L.~D.
  Jackel, M.~Monfort, U.~Muller, J.~Zhang, \emph{et~al.}, ``End to end learning
  for self-driving cars,'' \emph{arXiv preprint arXiv:1604.07316}, 2016.

\bibitem{McAfee}
``From machine learning to artificial intelligence,''
  \url{https://www.mcafee.com/enterprise/en-us/solutions/machine-learning.html},
  accessed: 2019-01-29.

\bibitem{8519598}
F.~Reway, W.~Huber, and E.~P. Ribeiro, ``Test methodology for vision-based adas
  algorithms with an automotive camera-in-the-loop,'' in \emph{2018 IEEE
  International Conference on Vehicular Electronics and Safety (ICVES)}, Sept.
  2018, pp. 1--7.

\bibitem{broadhurst2005monte}
A.~Broadhurst, S.~Baker, and T.~Kanade, ``Monte carlo road safety reasoning,''
  in \emph{2005 IEEE Intelligent Vehicles Symposium (IV)}.\hskip 1em plus 0.5em
  minus 0.4em\relax IEEE, June 2005, pp. 319--324.

\bibitem{lefevre2014survey}
S.~Lef{\`e}vre, D.~Vasquez, and C.~Laugier, ``A survey on motion prediction and
  risk assessment for intelligent vehicles,'' \emph{ROBOMECH journal}, vol.~1,
  no.~1, pp. 1--14, 2014.

\bibitem{ziegler2014trajectory}
J.~Ziegler, P.~Bender, T.~Dang, and C.~Stiller, ``Trajectory planning for
  bertha—a local, continuous method,'' in \emph{2014 IEEE Intelligent
  Vehicles Symposium Proceedings (IV)}.\hskip 1em plus 0.5em minus 0.4em\relax
  IEEE, June 2014, pp. 450--457.

\bibitem{ferguson2008motion}
D.~Ferguson, T.~M. Howard, and M.~Likhachev, ``Motion planning in urban
  environments,'' \emph{Journal of Field Robotics}, vol.~25, no. 11-12, pp.
  939--960, 2008.

\bibitem{gaussjordan}
R.~Ansorge and H.~J. Oberle, \emph{Mathematik für Ingenieure: Band 1: Lineare
  Algebra und analytische Geometrie, Differential- und Integralrechnung einer
  Variablen}.\hskip 1em plus 0.5em minus 0.4em\relax Wiley-VCH Verlag GmbH \&
  Co, 2000.

\bibitem{RAADeu}
FGSV, \emph{Richtlinien für die Anlage von Autobahnen (RAA)}.\hskip 1em plus
  0.5em minus 0.4em\relax Forschungsgesellschaft für Straßen- und
  Verkehrswesen e.V., 2008.

\bibitem{mitschke1972dynamik}
M.~Mitschke and H.~Wallentowitz, \emph{Dynamik der Kraftfahrzeuge}.\hskip 1em
  plus 0.5em minus 0.4em\relax Springer, 1972, vol.~4.

\bibitem{schramm2014vehicle}
D.~Schramm, M.~Hiller, and R.~Bardini, \emph{Vehicle Dynamics: Modeling and
  Simulation}.\hskip 1em plus 0.5em minus 0.4em\relax Springer, 2014.

\bibitem{iihs2009carsize}
{Insurance Institute for Highway Safety}, ``Special issue: Car size, weight and
  safety,'' \emph{Status Report}, vol.~44, no.~4, 2009.

\bibitem{cartype}
K.~Van~Miert, \emph{Case No COMP/M.1406-HYUNDAI/KIA}.\hskip 1em plus 0.5em
  minus 0.4em\relax Office for Official Publications of the European
  Communities, 1999.

\bibitem{vehicleinertia}
G.~Heydinger, R.~Bixel, W.~R. Garrott, M.~Pyne, J.~G. Howe, and D.~A. Guenther,
  \emph{Measured Vehicle Inertial Parameters-NHTSAs Data Through November
  1998}.\hskip 1em plus 0.5em minus 0.4em\relax Society of Automotive
  Engineers, 1999.

\bibitem{isermann2006fahrdynamik}
R.~Isermann, \emph{Fahrdynamik-Regelung: Modellbildung, Fahrerassistenzsysteme,
  Mechatronik}.\hskip 1em plus 0.5em minus 0.4em\relax Vieweg \& Sohn Verlag,
  2006.

\bibitem{driversteer}
G.~J. Forkenbrock and D.~Elsasser, \emph{An Assessment of Human Driver Steering
  Capability}.\hskip 1em plus 0.5em minus 0.4em\relax National Highway Traffic
  Safety Administration, 2005.

\bibitem{pedestrian1}
B.~A. Schlake, ``Mathematical models for pedestrian motion,'' Diplomarbeit,
  Westfälische Wilhelms-Universität Münster, 2008.

\bibitem{zkebala2012pedestrian}
J.~Z{\k{e}}bala, P.~Ci{\k{e}}pka, and A.~RezA, ``Pedestrian acceleration and
  speeds,'' \emph{Problems of Forensic Sciences}, vol.~91, pp. 227--234, 2012.

\bibitem{shimrat1962algorithm}
M.~Shimrat, ``Algorithm 112: Position of point relative to polygon,''
  \emph{Communications of the ACM}, vol.~5, no.~8, p. 434, 1962.

\bibitem{haines1994point}
E.~Haines, ``Point in polygon strategies,'' \emph{Graphics Gems IV}, vol. 994,
  pp. 24--26, 1994.

\bibitem{leissa2018anydsl}
R.~Leißa, K.~Boesche, S.~Hack, A.~Pérard-Gayot, R.~Membarth, P.~Slusallek,
  A.~Müller, and B.~Schmidt, ``{AnyDSL}: A partial evaluation framework for
  programming high-performance libraries,'' \emph{Proceedings of the ACM on
  Programming Languages (PACMPL)}, vol.~2, no. OOPSLA, pp. 119:1--119:30, Nov.
  2018.

\bibitem{leissa2015shallow}
R.~Leißa, K.~Boesche, S.~Hack, R.~Membarth, and P.~Slusallek, ``Shallow
  embedding of {DSLs} via online partial evaluation,'' in \emph{Proceedings of
  the International Conference on Generative Programming: Concepts \&
  Experiences (GPCE)}.\hskip 1em plus 0.5em minus 0.4em\relax ACM, 2015, pp.
  11--20.

\bibitem{iihs2004evidence}
{Insurance Institute for Highway Safety}, ``Evidence mounts that reducing force
  of airbag inflation lowers risk,'' \emph{Status Report}, vol.~39, no.~7,
  2004.

\bibitem{iihs2003headprotecting}
{Insurance Institute for Highway Safety}, ``Head-protecting side airbags reduce
  driver fatality risk by 45 percent,'' \emph{Status Report}, vol.~38, no.~8,
  2003.

\end{thebibliography}

\end{document}